\documentclass[12pt,preprint]{aastex}
\usepackage{lscape}
\usepackage{multirow}
\usepackage{booktabs}
\usepackage{graphicx}
\usepackage{amsmath}
\usepackage{longtable}
\usepackage{rotating}
\usepackage{CJKutf8}
\usepackage[caption=false]{subfig}
\usepackage{makecell}

\newcommand{\rsub}{R_{\rm sub}}
\newcommand{\mj}{M_{\rm J}}
\newcommand{\msun}{M_\odot}

\newcommand{\rc}{R_{\rm c}}

\begin{document}

\title{The Sizes and Depletions of the Dust and Gas Cavities in the Transitional Disk J160421.7-213028}

\shortauthors{Dong et al.}

\author{Ruobing Dong\altaffilmark{1,2}, Nienke van der Marel\altaffilmark{3}, Jun Hashimoto\altaffilmark{4}, Eugene Chiang\altaffilmark{2}, Eiji Akiyama\altaffilmark{5}, Hauyu Baobab Liu\altaffilmark{6}, Takayuki Muto\altaffilmark{7}, Gillian R. Knapp\altaffilmark{8}, Takashi Tsukagoshi\altaffilmark{9}, Joanna Brown\altaffilmark{10}, Simon Bruderer\altaffilmark{11}, Shin Koyamatsu\altaffilmark{12}, Tomoyuki Kudo\altaffilmark{13}, Nagayoshi Ohashi\altaffilmark{4}, Evan Rich\altaffilmark{14}, Mayama Satoshi\altaffilmark{15,16}, Michihiro Takami\altaffilmark{17}, John Wisniewski\altaffilmark{14}, Yi Yang\altaffilmark{18}, Zhaohuan Zhu\altaffilmark{19}, Motohide Tamura\altaffilmark{4,12}}

\altaffiltext{1}{Steward Observatory, University of Arizona, Tucson, AZ, 85721, rdong@email.arizona.edu}
\altaffiltext{2}{Department of Astronomy, University of California at Berkeley, 94720, Berkeley, CA}
\altaffiltext{3}{Institute for Astronomy, University of Hawaii, Honolulu, HI 96822, marel@hawaii.edu}
\altaffiltext{4}{Astrobiology Center, National Institutes of Natural Sciences, 2-21-1 Osawa, Mitaka, Tokyo 181-8588 Japan}
\altaffiltext{5}{National Astronomical Observatory of Japan, 2-21-1, Osawa, Mitaka, Tokyo, 181-8588, Japan}
\altaffiltext{6}{European Southern Observatory (ESO), Karl-Schwarzschild-Strasse 2, D-85748 Garching, Germany}
\altaffiltext{7}{Division of Liberal Arts, Kogakuin University, 1-24-2 Nishi-Shinjuku, Shinjuku-ku, Tokyo 163-8677}
\altaffiltext{8}{Department of Astrophysical Sciences, Princeton University, Princeton, NJ 08544}
\altaffiltext{9}{College of Science, Ibaraki University, Bunkyo 2-1-1, Mito, Ibaraki 310-8512, Japan}
\altaffiltext{10}{Boston Fusion}
\altaffiltext{11}{Max-Planck-Institut for Extraterrestrische Physik, Giessenbachstrasse 1, 85748 Garching, Germany}
\altaffiltext{12}{Department of Astronomy, Graduate School of Science, The University of Tokyo, Hongo 7-3-1, Bunkyo-ku, Tokyo 113-0033, Japan}
\altaffiltext{13}{Subaru Telescope, National Astronomical Observatory of Japan, 650 North Aohoku Place, Hilo, HI 96720, USA}
\altaffiltext{14}{Homer L. Dodge Department of Physics, University of Oklahoma, Norman, OK 73071, USA}
\altaffiltext{15}{The Center for the Promotion of Integrated Sciences, The Graduate University for Advanced Studies~(SOKENDAI), Shonan International Village, Hayama-cho, Miura-gun, Kanagawa 240-0193, Japan}
\altaffiltext{16}{Department of Astronomical Science, The Graduate University for Advanced Studies~(SOKENDAI), 2-21-1 Osawa, Mitaka, Tokyo 181-8588, Japan}
\altaffiltext{17}{Institute of Astronomy and Astrophysics, Academia Sinica, PO Box 23-141, Taipei 10617, Taiwan, ROC}
\altaffiltext{18}{Department of Astronomical Science, The Graduate University for Advanced Studies, 2-21-1 Osawa, Mitaka, Tokyo 181-8588, Japan}
\altaffiltext{19}{Department of Physics and Astronomy, University of Nevada, Las Vegas, 4505 South Maryland Parkway, Las Vegas, NV 89154}

\clearpage

\begin{abstract}
  We report ALMA Cycle 2 observations of 230 GHz (1.3 mm) dust
  continuum emission, and $^{12}$CO, $^{13}$CO, and C$^{18}$O J = 2-1 line
  emission, from the Upper Scorpius transitional disk [PZ99]
  J160421.7-213028, with an angular resolution of
  $\sim$$0\arcsec.25$ (35 AU).
  Armed with these data and existing H-band scattered
  light observations, we measure the size and depth of the
  disk's central cavity, and the sharpness of its outer edge,
  in three components: sub-$\mu$m-sized ``small''
  dust traced by scattered light, millimeter-sized ``big'' dust
  traced by the millimeter continuum, and gas traced by
  line emission. Both dust
  populations feature a cavity of radius $\sim$70 AU that is depleted by factors of at least 1000 relative to the dust density just outside. The millimeter continuum data are well explained by a cavity with a sharp edge. Scattered light observations can be fitted with a cavity in small dust that has 
  either a sharp edge at 60 AU, or an edge that transitions
  smoothly over an annular width of 10 AU near 60 AU.
  In gas,
  the data are consistent with a cavity that is smaller,
  about 15 AU in radius,
  and whose surface density at 15 AU is $10^{3\pm1}$
  times smaller than the surface density at 70 AU;
  the gas density grades smoothly between these two radii.
  The CO isotopologue observations rule out a sharp drop in gas surface
  density at 30 AU or a double-drop model as found by previous
  modeling. Future
  observations are needed to assess the nature of these
  gas and dust cavities, e.g., whether they are opened
  by multiple as-yet-unseen planets or photoevaporation.

\end{abstract}



\section{Introduction}\label{sec:intro}

Transitional disks are gaseous protoplanetary disks with a central
depleted region\footnote{In the literature the central depleted region
  has been called a ``gap'' or a ``cavity'',
  depending on whether the structure extends all the way to the
  star. In this paper we refer to the structure in J1604 as a cavity.} \citep[see
the review by][]{espaillat14}. They mark a crucial phase in disk
evolution, intermediate between fully gas-rich and gas-depleted systems. Their
existence was first suggested by the distinctive near-to-mid-infrared
(NIR-MIR) dips in their spectral energy distributions (SEDs;
e.g. \citealt{storm89,skrutskie90};
\citealt{calvet05,espaillat07,espaillat08}), and later confirmed by
resolved images in
NIR scattered light \citep[e.g.,][]{thalmann10,
  hashimoto12, mayama12, garufi13, quanz13gap, avenhaus14hd100546, avenhaus14,
  tsukagoshi14} and by resolved mm-wave maps
of dust continuum and gas
line emission \citep[e.g.][]{andrews11, mathews12,
  tang12, isella13, vandermarel13, fukagawa13, vandermarel14, perez14,
  zhang14, vandermarel15irs48, canovas15sz91, hashimoto15}.

{\it What opens the cavities in transitional disks?} This is
still an open question. The leading hypothesis is dynamical
sculpting by planets (or more massive companions) inside the cavity. Cavity
opening is a natural outcome of tidal interactions between a disk and
companions \citep[e.g.,][]{lin93, artymowicz94, bryden99,
  kley12}. While a single gap opened by one giant planet may be too
narrow to account for the observed cavity sizes in the gas and scattered
light, \citet{zhu11}, \citet{dodsonrobinson11}, and \citet{dong15gaps}
explored the possibility of opening a cavity by multiple
giant planets (see also \citealt{duffell15dong}). In this scenario, the
sharpness of the gas cavity edge increases with planet mass
\citep[e.g.,][]{duffell15gap}. Large gradients in gas surface
density can
cause the appearance of the cavity (e.g., its size) to depend on
wavelength. Because mm-sized dust particles can pile up at the
pressure bump outside the gas cavity edge \citep[this is called the
``dust filtration'' effect;][]{rice06, zhu12, pinilla12-dusttrapping,
  pinilla12-diffcavsize, dejuanovelar13}, cavities viewed in the mm
continuum can be larger than they appear in scattered light and gas
observations.  Another consequence of cavity opening by companions is a
reduced accretion rate onto the star, depending on how much of the
disk accretion flow is diverted onto the companions. A small inner
disk may remain if no companions are present there.

The main alternative non-planet mechanism for clearing big cavities in
transitional disks is photoevaporation \citep[e.g.,][]{clarke01,
  owen10, owen11, suzuki10}. In this scenario, stellar radiation ionizes surface
layers of the disk and launches a wind from the outer disk; if the disk accretion
rate is smaller than the wind mass loss rate, the inner
disk is starved and a cavity opens.
In this scenario, the cavity edge in both gas and dust tends to be sharp
\citep[e.g.,][]{alexander06, alexander07}, and since the disk is cleared from the inside
out, the accretion on to the star is expected to be very low or zero
\citep[e.g.,][]{owen11}.
Particle trapping at the gap edge can also occur in photoevaporated cavities.

Other mechanisms for explaining large cavities in observations have also been
proposed, such as grain growth \citep[e.g.,][]{birnstiel12} and disk
shadowing \citep[e.g.,][for cavity/ring structures seen in scattered
light]{garufi14}. However, these mechanisms cannot reproduce
certain observed features in the disks such as cavity edges \citep{birnstiel12, dong15shadow}.

Identifying the origin of the cavity has important implications for disk evolution and planet formation. Multi-wavelength, spatially resolved observations are needed, as various cavity formation mechanisms predict different structures for different components, resulting in different observed disk morphologies at different wavelengths. [PZ99] J160421.7-213028 (hereafter J1604), a transitional
disk heavily scrutinized in recent years, provides an excellent case study. This nearly face-on (inclination $\sim6^\circ$; \citealt{mathews12}) system is located at $\sim$145~pc in the $\sim$5--10~Myr old Upper Scorpius star forming region \citep{dezeeuw99, pecaut12}. The central source is a pre-main-sequence star with a spectral type of K2, an effective temperature of $T_{\rm eff}\sim4500$K, and a mass $M_\star\sim1\msun$ \citep{dahm09, mathews12, carpenter14}. Its cavity is one of the largest, extending to $\sim$70 AU, as vividly revealed in NIR polarized light by Subaru/HiCIAO \citep[$H$-band;][]{mayama12} and VLT/SPHERE \citep[$R^\prime$-band;][]{pinilla15j1604}. Millimeter observations using SMA \citep[0.88 mm;][]{mathews12} and ALMA \citep[cycle0, 0.88~mm, band 7;][]{zhang14} have resolved the cavity in dust and CO $J=$3-2 emission, with angular resolutions of $0\arcsec.51\times0\arcsec.34$ and $0\arcsec.73\times0\arcsec.46$, respectively. As a transitional disk, J1604 has several peculiar properties. In particular, \citet{owen16} pointed out that most transitional disks can be classified into two classes: one with small holes ($\lesssim$ 10 AU) and low accretion rates ($<10^{-9}M_\odot$ yr$^{-1}$), and another with large holes ($\gtrsim$ 20 AU) and high accretion rates $\sim10^{-8}M_\odot$ yr$^{-1}$. J1604 belongs to neither: it has one of the largest holes, and yet it is hardly accreting \citep{mathews12}.

In this paper, we present new ALMA Cycle 2 Band-6 (1.3 mm)
dust
continuum and $J=2$--1 line observations
for three CO isotopologues
($^{12}$CO/$^{13}$CO/C$^{18}$O), with an angular
resolution of $\sim$$0\arcsec.25$, the highest at mm
wavelengths to date.
These data, in combination with a well-sampled SED and
resolved observations at 0.6~$\micron$, 1.6~$\micron$, and
0.88 mm, afford an unprecedentedly
detailed examination of a transitional disk.
We probe cavity structures in dust and gas using parametrized
axisymmetric disk models and dust and line radiative
transfer simulations (Section~\ref{sec:method}),
to answer three basic questions 
(Section~\ref{sec:properties}): 
\begin{enumerate}
\item What are the sizes of the cavities seen in various disk components: ``small'' sub-$\mu$m-sized dust traced by
scattered light, ``big'' millimeter-sized grains traced
by mm dust continuum emission, and gas traced by CO?
\item How depleted are the cavities in the various disk components?
\item How sharp are the cavity edges in the various disk components?
\end{enumerate}
A summary and discussion are given at the end (Section 5).


\section{ALMA Observations and Data Reduction}\label{sec:observations}

J1604 \citep[RA 16:04:21.643, Dec -21:30:28.72;][]{cutri13}
was observed with the Atacama Large Millimeter/submillimeter Array (ALMA) in Band 6 (230 GHz) during ALMA Cycle 2 observations (program ID: 2013.1.01020.S, PI: T.~Tsukagoshi) in July 2015. The observations were conducted in four spectral windows: two with bandwidths of 117.19~MHz (and channel widths of 61.035~kHz; equivalent to a velocity resolution of $\sim$0.08~km~s$^{-1}$) centered on $^{12}$CO~(2--1) and $^{13}$CO~(2--1); one with a bandwidth of 468.75~MHz (and a channel width of 0.244~MHz; equivalent to a velocity resolution $\sim$0.33~km~s$^{-1}$) centered on C$^{18}$O~(2--1); and a fourth spectral window for continuum observations with a higher sensitivity bandwidth of 1875.00~MHz (and a channel width of 31.250~MHz). The flux and bandpass were calibrated with the quasar J1517-243, which was used as a bandpass calibrator as well. The gain/phase calibrator was quasar J1559-2442. The total on-source integration time was 316 seconds.
  
The data were calibrated with CASA \citep[version 4.2]{mcmullin07} following the calibration scripts provided by EA-ARC, and then imaged in CASA using the CLEAN algorithm \citep{rau11}. The continuum data were concatenated from four spectral windows providing $\sim$2.6~GHz of continuum bandwidth. 

The continuum data were cleaned using Briggs weighting with a robust factor of 0.5, and the line data were cleaned using natural weighting, resulting in a beam size of $\sim0\arcsec.25$.
Natural weighting was chosen over Briggs weighting for the line data for better image recovery as the signal-to-noise ratio is lower for the line data. 

The 230 GHz continuum emission and the three CO 2--1 isotopologues $^{12}$CO (230.538 GHz), $^{13}$CO (220.398677 GHz) and C$^{18}$O (219.56036 GHz) were all imaged. Table \ref{tab:obsproperties} summarizes the continuum and line data. 

\begin{figure}
\includegraphics[width=\textwidth]{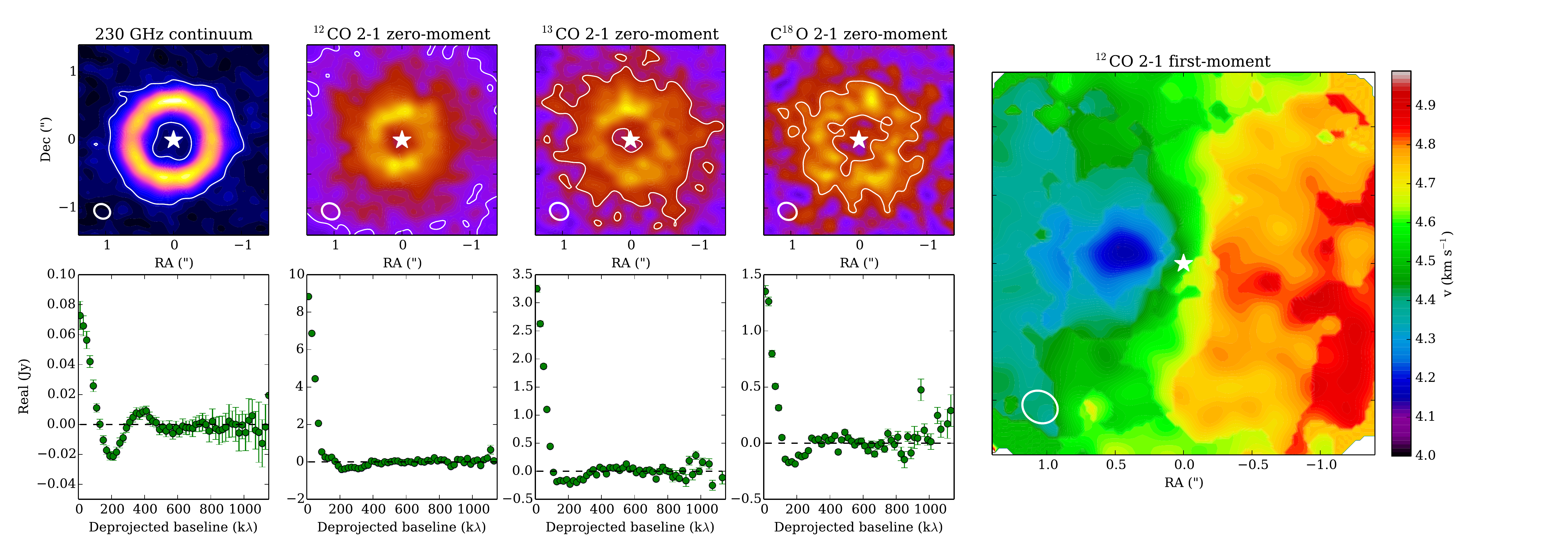}
\caption{ALMA observations of the
230 GHz continuum, $^{12}$CO, $^{13}$CO and C$^{18}$O $J=$2--1 moment maps and visibility curves of J1604-2130. The zero-moment map is the total intensity, and the first-moment map is the velocity field. The 3$\sigma$ contours are given in white, with $\sigma_{\rm continuum}$=0.11 mJy and  $\sigma_{\rm line}$=11 mJy km s$^{-1}$ for the integrated line emission.}
\label{fig:observations}
\end{figure}

\begin{figure}
\includegraphics[width=\textwidth]{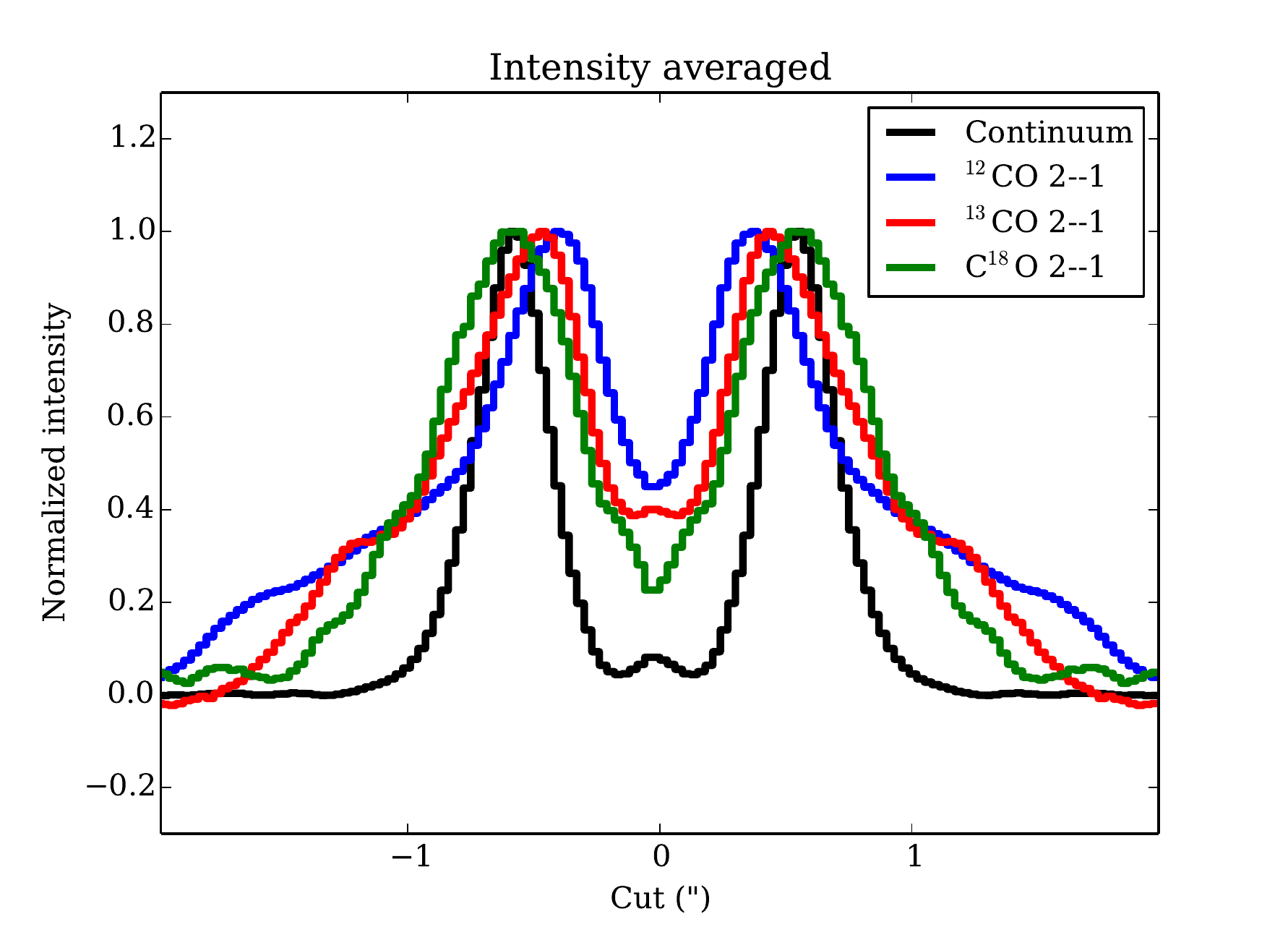}
\caption{Azimuthally averaged normalized intensity of the ALMA observations of the 230 GHz continuum, $^{12}$CO, $^{13}$CO and C$^{18}$O $J=$2--1 zero-moment maps of J1604-2130 from Figure \ref{fig:observations}.}
\label{fig:observationsavg}
\end{figure}

Figure \ref{fig:observations} shows the continuum map, the zero-moment maps (total line intensity) for all three CO lines, and the first-moment map (the velocity field) in $^{12}$CO 2--1. The continuum is detected with a peak signal-to-noise ratio of 36 ($\sigma$=0.11 mJy beam$^{-1}$), the integrated line intensities have a peak signal-to-noise ratio of 17, 9 and 6 for $^{12}$CO, $^{13}$CO and C$^{18}$O, respectively, with $\sigma_{\rm line}\approx$11 mJy km s$^{-1}$ for the integrated emission. The $\sigma$ is determined by measuring the standard deviation in a ring outside a 2" radius in the continuum and zero-moment maps. The first-moment map is consistent with Keplerian rotation, and the stellar position derived from the first-moment map is RA 16$^h$04$^m$21.638$^s$, Dec -21$^{\circ}$30'28.98", consistent with the position of the star in the optical/IR. We derive a position angle of 80$^\circ$ and an inclination of 6$^{\circ}$, consistent with previous estimates based on ALMA Cycle 0 data. The 230 GHz continuum image shows a narrow, azimuthally symmetric ring, as was found in previous observations at 345 GHz with lower spatial resolution \citep[][]{mathews12,zhang14,vandermarel15}. The zero moment maps of the CO lines show rings as well, but with smaller inner radii than the continuum ring, again consistent with previous findings. Figure \ref{fig:observationsavg} shows the azimuthally averaged radial cuts for continuum and integrated line emission.
The inner radii of the $^{13}$CO and C$^{18}$O rings appear slightly larger than that of the $^{12}$CO ring; this may be an effect of their optical depths differing
according to their different abundances.
The azimuthally averaged (after correcting for the small inclination) visibility profiles of both continuum and integrated CO data (bottom panels of Figure \ref{fig:observations}) are consistent with ring profiles as well: all profiles show clear nulls, at $\sim$130 k$\lambda$ (continuum), $\sim$170 k$\lambda$ ($^{12}$CO), $\sim$110 k$\lambda$ ($^{13}$CO) and $\sim$110 k$\lambda$ (C$^{18}$O). 

Non-Keplerian motion may indicate the presence of fast radial flows or disk warps \citep[e.g.,][]{rosenfeld14, casassus15}, or turbulence caused by various instabilities \citep[e.g.,][]{simon15, flaherty15}. The ALMA observations of J1604 do not show any clear indications of non-Keplerian motions, but the nearly face-on orientation of the disk makes velocity determinations difficult.


\section{Modeling}\label{sec:method}

Protoplanetary disks contain gas and variously sized dust grains. Dust dominates the opacity at nearly all continuum wavelengths.
For the purpose of modeling observations, a disk may be approximated as a three-component system, each primarily responsible for observations in one wavelength range \citep[e.g.,][]{dong12cavity, vandermarel15}: (1) gas --- vertically supported by pressure, producing CO emission; (2) sub-micron-sized dust (hereafter ``small'' dust) --- generally well-mixed with gas in the vertical direction and mainly responsible for the NIR scattered light; and (3) $\sim$mm-sized grains (hereafter ``big'' dust) --- possibly concentrated in regions of high gas pressure including the disk midplane \citep{dullemond04dustsettling, dalessio06, birnstiel10}, and mainly responsible for mm continuum emission. The distribution of small dust grains does affect the mm continuum by regulating the disk temperature (starlight is intercepted and reprocessed first by small dust at large altitude); however, this dependence of the mm-wave map on small dust is relatively minor: midplane temperatures at a given radius vary by a factor of a few at most using different dust distributions, while surface densities in all three disk components vary by several orders of magnitudes across the cavity, as we will show.

We use radiative transfer simulations and parametrized disk models to produce synthetic observations and compare them with the data. The models are axisymmetric with as few radial parameters as needed to match the observations. We do not aim at formally fitting the observations in a $\chi^2$ manner, as this is impractical given the large number of degrees of freedom; fitting is done by eye instead. We are interested in obtaining rough estimates for basic properties of the cavity (as viewed in each component): the cavity size, degree of depletion, and the sharpness of its edge.
We employ two radiative transfer tools to produce synthetic observations. For scattered light, we use the \citet{whitney13} Monte Carlo radiative transfer (MCRT) code; for dust continuum and CO line emission, we use the physical-chemical DALI code \citep{bruderer12, bruderer13}. We treat the small dust separately from the big dust and the gas in the modeling, and largely follow the procedures described by \citet{dong12pds70} for scattered light and \citet{vandermarel16} for mm continuum and line emission.

The disk starts from the dust sublimation radius $\rsub$, corresponding to a temperature of $\sim$1500 K ($\sim$0.055 AU for J1604), and extends to an outer radius $R_{\rm out}$. For the central source, we use a pre-main sequence star of spectral type K2, radius 1.4 $R_\odot$, mass 1.0 $M_\odot$, and temperature 4500~K. The star is not known to be accreting \citep[$\dot{M}<10^{-11} M_\odot$/yr;][]{mathews12}. Our model's surface density profile $\Sigma(R)$ divides into an outer disk and a depleted inner disk for all three components, as illustrated in Figure~\ref{fig:fiducial}:
\begin{equation}
 \Sigma(R) =
  \begin{cases}
   \delta_{\rm cav}(R)\Sigma_{\rm 0}\left(\frac{\rc}{R}\right)^\gamma e^{-R/\rc}, R\leq R_{\rm cav} & \text{Cavity} \\
   \ \ \ \ \ \ \ \ \ \Sigma_{\rm 0}\left(\frac{\rc}{R}\right)^\gamma e^{-R/\rc}, R_{\rm cav}<R\leq R_{\rm out} & \text{Outer Disk}
  \end{cases}
\label{eq:sigmacavity}
\end{equation}
where the exponential length scale $\rc$, power-law index $\gamma$, cavity depletion
factor $\delta_{\rm cav}(R)$, and cavity radius $R_{\rm cav}$ are parameters specific to each of the three
disk components (small dust, big dust, and gas).
For small dust we introduce an additional rim structure from $\rsub$ to $R_{\rm rim}$ to account for possible NIR excess:\footnote{J1604 has been labeled a possible variable source by \citet{dahm09}; the IRAC data indicate no NIR excess, while the Spitzer IRS spectrum indicate a NIR excess. The later WISE data at 3.4 and 4.6 $\mu$m \citep{cutri12} are consistent with the IRS spectrum but not the IRAC photometry. We adopt the WISE and IRS data in this paper (the IRAC data are not plotted in Figure~\ref{fig:fiducial}).} the surface density of small dust inside the rim is given by $\delta_{\rm rim}\Sigma_{\rm 0}\left(\frac{\rc}{R}\right)^\gamma e^{-R/\rc}$.
Note that $\delta_{\rm cav}$ can vary with radius. In some of our models
we will set $\delta_{\rm cav}$ to be constant, while in others we will allow
it vary with radius to introduce additional structure.

In the vertical direction $z$, strongly irradiated (i.e., passive) protoplanetary disks are roughly isothermal, except in the tenuous upper layers \citep{chiang97, dullemond02}. In hydrostatic equilibrium, the vertical gas density follows a Gaussian profile:
\begin{equation}
\rho(R,z)=\frac{\Sigma(R)}{\sqrt{2\pi}h} e^{-z^2/2h^2} \,,
\label{eq:rhorz}
\end{equation}
where $h$ is the scale height. The big grains tend to settle to the midplane; we assume their vertical distribution also obeys
a Gaussian but with a smaller $h$. Radially, the scale height is assumed to vary with radius as
\begin{equation}
h\propto R^\psi \,,
\label{eq:h}
\end{equation}
where $\psi$ is a component-dependent constant. 

We adopt the interstellar medium dust model of
\citet{kim94} for small dust
(composed of silicate, graphite, and amorphous carbon) with a size distribution that runs from $\sim$$0.002\micron$ to $\sim$$0.25\micron$. As J1604 is nearly face-on, the scattering angle everywhere in the scattered light image is close to $90^\circ$. We assume the \citet{andrews11} big dust model for our big dust, which has a minimum size of 0.005 $\micron$ and a maximum size of 1 mm with a power law differential size ($s$) distribution $n(s)\propto s^{-3.5}$. Mie scattering is assumed for both dust populations.

For the modeling of the CO isotopologues, the DALI code \citep{bruderer12,bruderer13} is used. DALI is a physical-chemical modeling code which solves the heating-cooling balance of the gas and chemistry simultaneously to determine the gas temperature, molecular abundances and molecular excitation in every position in the disk for a given density structure. DALI uses a chemical reaction network of about 110 species and 1500 reactions, including basic grain-surface reactions (freeze-out, sublimation and hydrogenation). DALI is required for proper interpretation of CO emission for several reasons: the gas and dust temperature are decoupled in disks, especially inside and at the cavity edges; the local CO abundance w.r.t. H$_2$ is lowered due to photodissociation and freeze-out and is thus not a direct gas density tracer; CO is formed and destroyed through various chemical reactions depending on the local conditions in the disk. DALI has been used to interpret several transition disks in spatially resolved CO observations \citep{bruderer14, vandermarel15, vandermarel16}. The full details on the DALI model are discussed in these papers as well. The assumed abundance ratios of the CO isotopologues in DALI are $^{12}$CO/$^{13}$CO=77 and $^{12}$CO/C$^{18}$O=560. The effects of isotope-selective photodissociation \citep[e.g.,][]{miotello14} have been checked but these do not significantly change the emission for our fiducial model.

In total, there are 23 parameters: $\Sigma_{\rm 0}$ (equivalent to the total disk mass), $R_{\rm c}$, $R_{\rm out}$, $\gamma$, $\psi$, $R_{\rm cav}$, and $\delta_{\rm cav}$ for each of the 3 components, plus $\delta_{\rm rim}$ and $R_{\rm rim}$ for the small dust. We use subscripts ``gas,'' ``small-dust,'' and ``big-dust'' to indicate each component.
We are mainly interested in the cavity size, depletion, and edge structure for each of the three components.
These parameters largely determine the cavity morphology in observations, while experiments have shown that our data are insensitive to many of the other parameters \citep{dong12cavity, dong12pds70, vandermarel15, vandermarel16}.

From our model we generate the SED, and images in $H$-band polarized light, mm continuum, and $^{12}$CO/$^{13}$CO/C$^{18}$O $J=$2--1 emission. For scattered light we use the Subaru/HiCIAO image by \citealt{mayama12} (data taken as part of the SEEDS planet and disk survey; \citealt{tamura09}), and for continuum and line emission we use the ALMA Cycle 2 data presented in this paper. For the $H$-band images, we convolve the full resolution model images with the observed HiCIAO point spread function to achieve the appropriate angular resolution. Synthetic ALMA images are convolved with the ALMA angular resolution as given in Table \ref{tab:obsproperties}. Also, we calculate the visibility profiles directly from the integrated gas moment maps and continuum images and compare these with the observed visibility profiles.


\section{Disk Properties}\label{sec:properties}

In this section, we first present a fiducial model that fits all the  observations reasonably well (Section~\ref{sec:fiducial}). We then vary the sizes (\ref{sec:size}), depletion factors (\ref{sec:depletion}), and sharpnesses of the cavity edges (\ref{sec:edge}) to explore the uncertainties.

\subsection{The Fiducial Model}\label{sec:fiducial}
 
\begin{figure}
\begin{center}
\includegraphics[width=\textwidth]{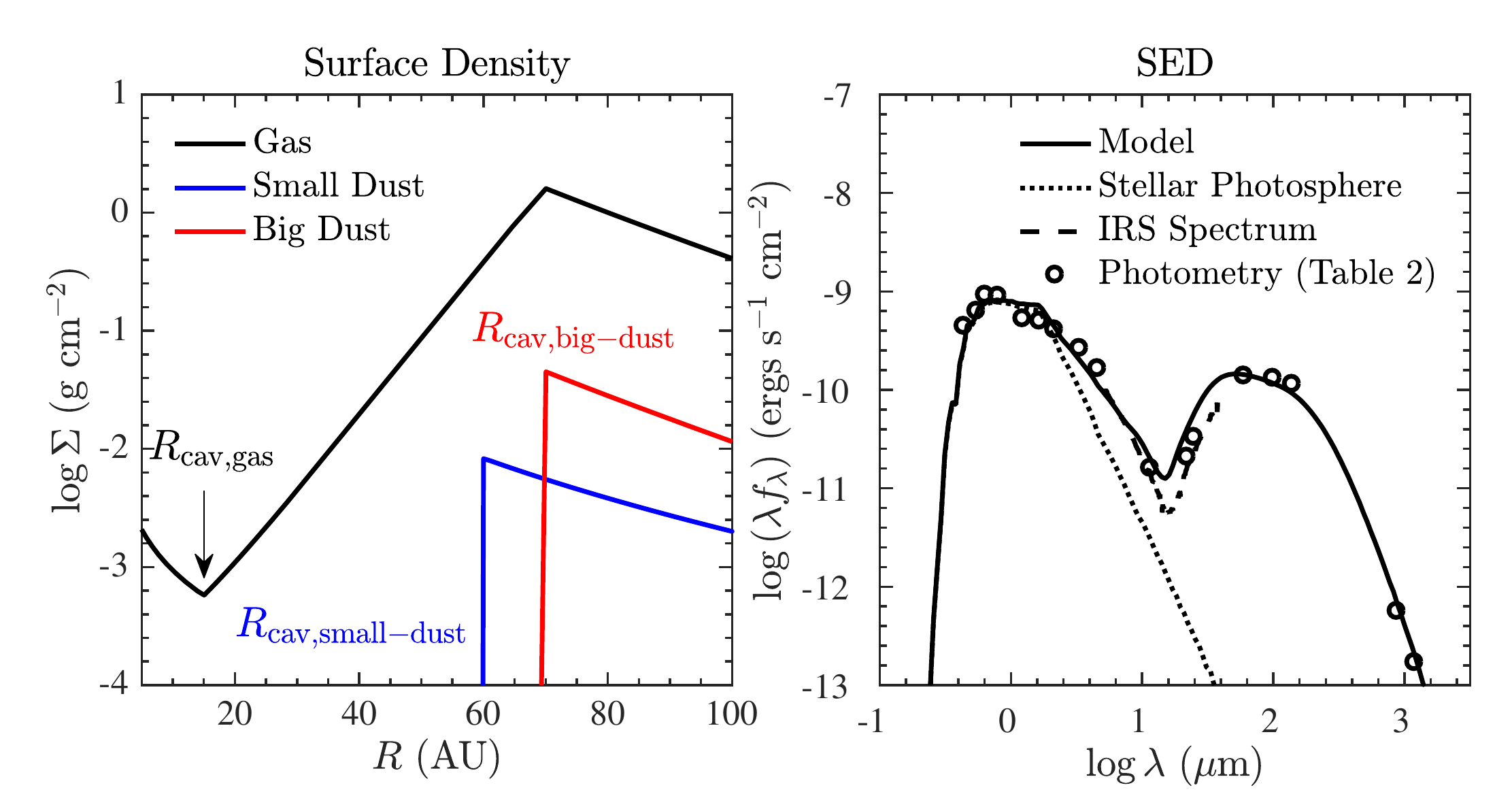}
\caption{Surface density of each fiducial model component and the SED of the fiducial model. Observational data for the SED are listed in Table~\ref{tab:sed} (IRAC data points are ignored, see Section~\ref{sec:fiducial} for details).}
\label{fig:fiducial}
\end{center}
\end{figure}

\begin{figure}
\begin{center}
\includegraphics[width=\textwidth]{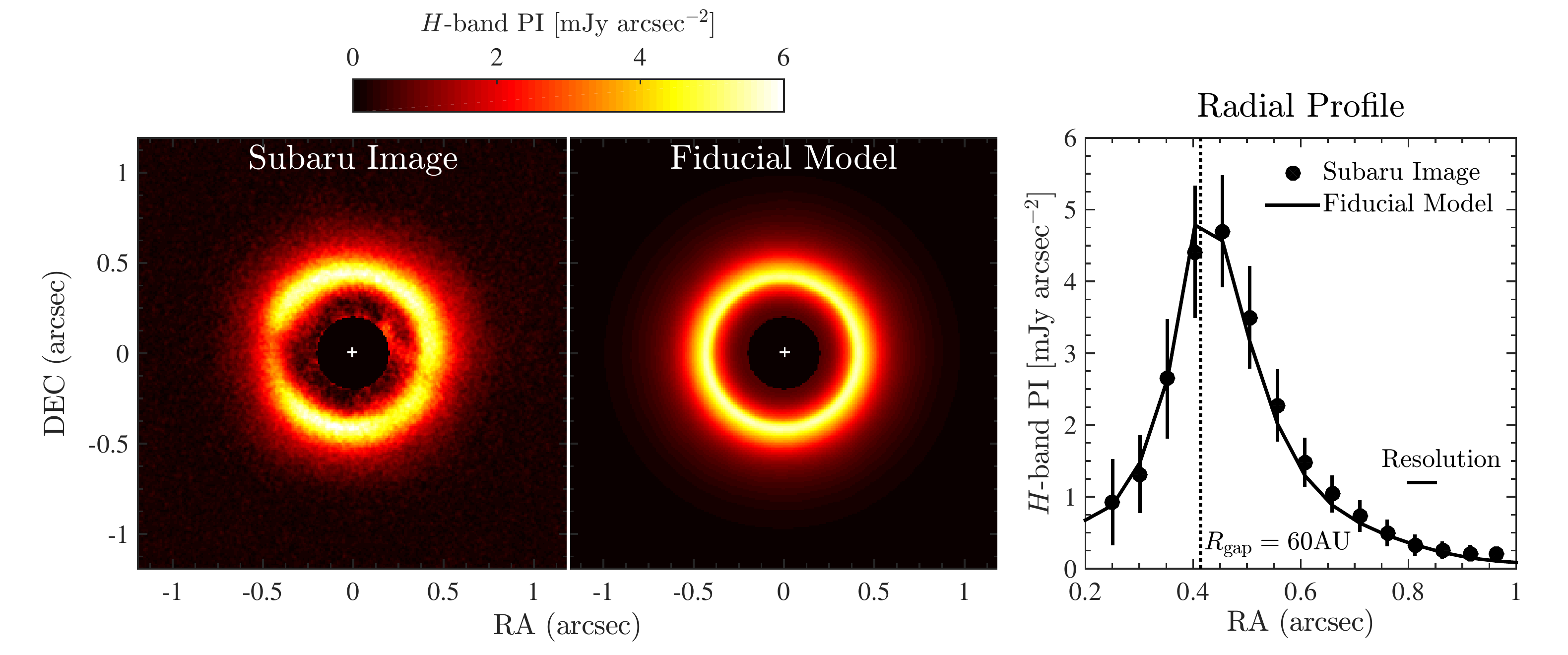}
\caption{$H$-band polarized intensity comparison between the fiducial model and the observations \citep{mayama12}. The error bars in the radial profile represent the root mean square scatter of the pixels in each annulus in the Subaru image. The horizontal line segment in the right panel indicates the angular resolution in the Subaru observation.}
\label{fig:nir_fiducial}
\end{center}
\end{figure}

\begin{figure}
\includegraphics[width=\textwidth]{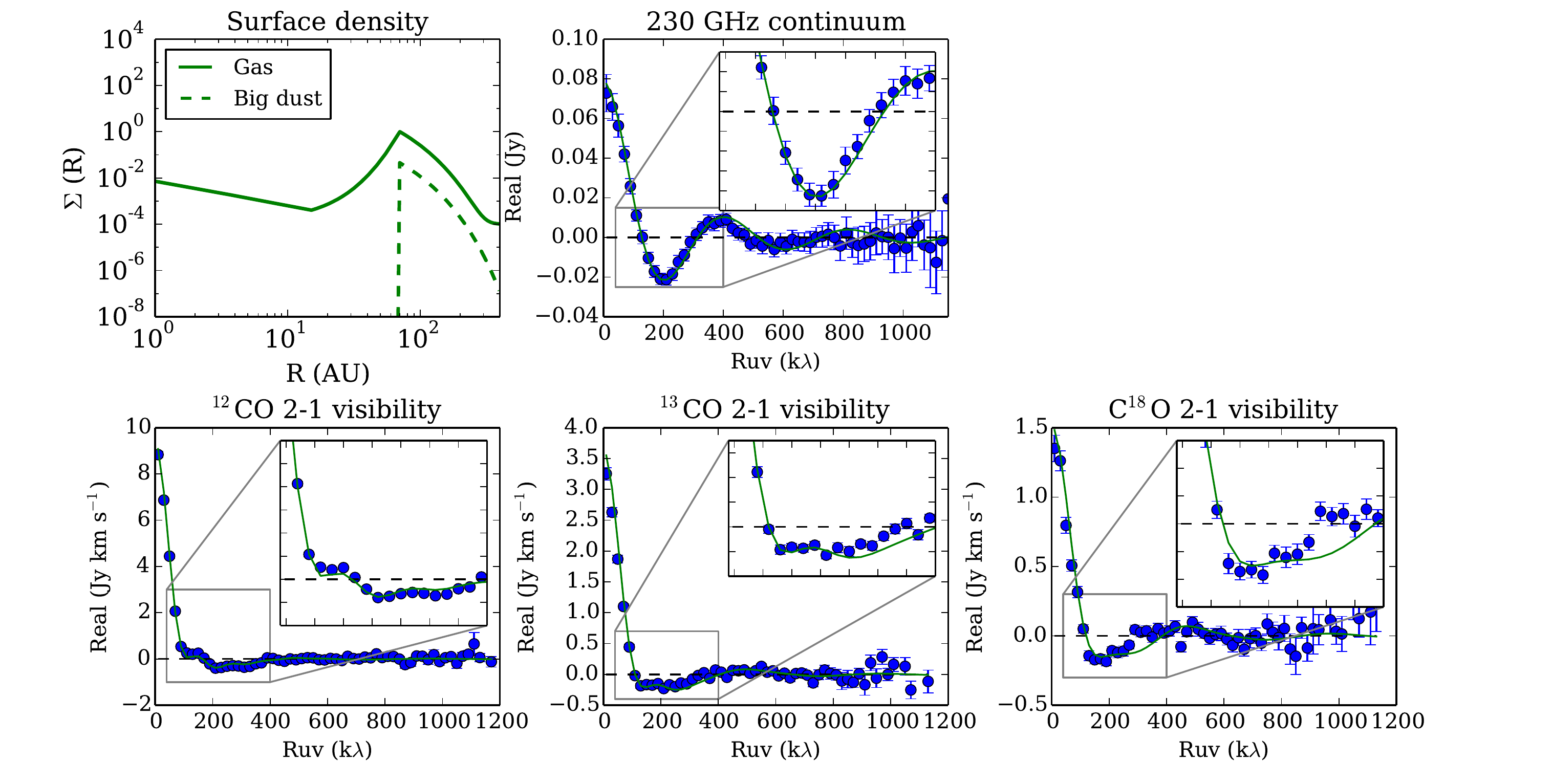}
\caption{Visibilities of the fiducial model, compared with the observations. All visibility data have been binned to 20 k$\lambda$ and deprojected. The data are shown in blue circles with corresponding error bars and the model as green lines. A black dashed line indicates the null line. An inset shows a blow-up of the profile around the location of the null. {\bf Top left:} The surface density profile of gas (solid) and big dust (dashed). {\bf Top middle:} Visibility profile of the 230 GHz continuum. {\bf Bottom left:} Integrated $^{12}$CO 2-1 visibility profile.  {\bf Bottom middle:} Integrated $^{13}$CO 2-1 visibility profile.  {\bf Bottom right:} Integrated C$^{18}$O 2-1 visibility profile. The model with the smooth density drop fits the data properly.
}
\label{fig:co-finalmodel}
\end{figure}

\begin{figure}
\includegraphics[width=\textwidth]{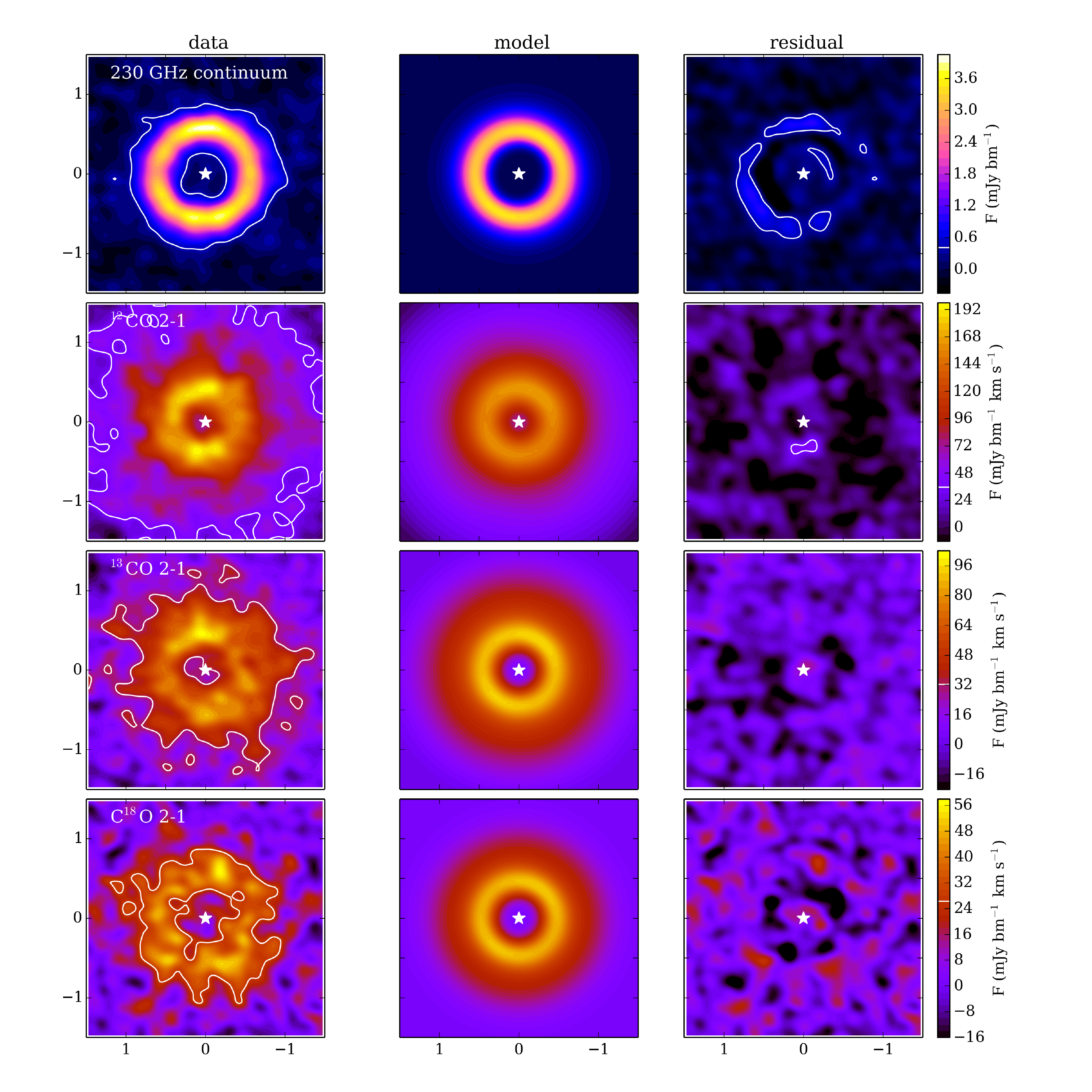}
\caption{The fiducial model compared with the observations at mm wavelengths. Top to bottom shows the 230 GHz continuum, and the zero-moment maps of $^{12}$CO 2-1, $^{13}$CO 2-1, and C$^{18}$O 2-1. Each row shows (from left to right) the data, the model and the residuals (produced by inversely transform the data$-$model residuals in the Fourier space back to the image space) on the same color scale. The 3$\sigma$ contours are given in white, with $\sigma_{\rm continuum}$=0.11 mJy and  $\sigma_{\rm line}\approx$11 mJy km s$^{-1}$ for the integrated line emission.}
\label{fig:residualmaps}
\end{figure}

Table~\ref{tab:base} lists the parameters of the fiducial model as portrayed in Figures~\ref{fig:fiducial}--\ref{fig:residualmaps}. Figure~\ref{fig:fiducial} shows the model surface density radial profiles for the three components and compares the global SED of the model with observations. Figure~\ref{fig:nir_fiducial} compares the H-band polarized intensity images;
Figure~\ref{fig:co-finalmodel} the visibilities of the models and data for the line emission and mm continuum;
and Figure~\ref{fig:residualmaps} the model and observed maps for the same.
The photometry and the {\it IRS} spectrum used to construct the SED are listed in Table~\ref{tab:sed}. The fiducial model qualitatively reproduces the SED, the image morphology at each wavelength, the radial profile of the scattered light, the mm visibilities, and the CO spectrum (none of the synthetic observations has been rescaled in flux). The fiducial model is also consistent with ALMA Cycle 0 345 GHz continuum and $^{12}$CO $J$=3-2 data (not shown).

The total dust mass in the model is 0.066 $\mj$ (0.013~$\mj$ in small dust and 0.053 $\mj$ in big dust, and the total gas mass is 2.5 $\mj$, resulting in a global gas-to-dust-mass ratio of $\sim$38:1. The scattered light and dust continuum observations are consistent with the simplest model --- an outer disk, a cavity that is completely empty (except possibly for a $<0.1$ AU inner disk in small dust, see below), and a sharp cavity edge. For the small dust we have an inner rim extending from $\rsub=0.055$~AU to $R_{\rm rim}=$0.07~AU, included to account for the occasional NIR excess (see footnote 2). Note that detailed SED fitting is beyond the scope of this paper. In reality grain sizes can vary across the disk, and relaxing our assumption of a single grain size distribution can help on the SED fitting. This inner rim does not affect the three resolved observations discussed here. The cavity sizes in the two dust populations differ slightly ($R_{\rm cav,small-dust}=60$~AU while $R_{\rm cav,big-dust}=70$~AU); however, we will see later that the difference is insignificant. For the gas, the simple cavity model --- a gas cavity of 30 AU with a sharp edge, used by \citet{vandermarel15} to fit the lower resolution Cycle 0 $^{12}$CO data --- does not fit the new ALMA $^{13}$CO and C$^{18}$O data (Section~\ref{sec:edge}). In order to fit all three isotopologues simultaneously, a smooth rather than a sharp cutoff at the gas cavity edge is required. 
We therefore add to the fiducial model for gas by
introducing a smooth exponential drop-off
in surface density between the big-dust cavity radius
$R_{\rm cav,big-dust}=70$~AU and the gas cavity
radius $R_{\rm cav,gas}=15$~AU. A full description of the gas surface density in the fiducial model is:
\begin{equation}
 \Sigma_{\rm gas}(R) =
  \begin{cases}
   \delta_{\rm cav,gas}(R)\Sigma_{\rm 0,gas}\left(\frac{R_{\rm c,gas}}{R}\right)^{\gamma_{\rm gas}} e^{-R/R_{\rm c,gas}} & R\leq R_{\rm cav,gas}\ \text{Cavity} \\
    \Sigma_{\rm gas}(R_{\rm cav,big-dust})\cdot e^{(R-R_{\rm cav,big-dust})/w} & \ R_{\rm cav,gas}\leq R\leq R_{\rm cav, big-dust}\ \text{Transition Region}\\
    \Sigma_{\rm 0,gas}\left(\frac{R_{\rm c,gas}}{R}\right)^{\gamma_{\rm gas}} e^{-R/R_{\rm c,gas}} & R_{\rm cav,big-dust}\leq R\leq R_{\rm out,gas}\ \text{Outer Disk}
  \end{cases}
\label{eq:sigmagas}
\end{equation}
where $w$ is
\begin{equation}
    w=\frac{R_{\rm cav,big-dust}-R_{\rm cav,gas}}{\ln[\Sigma_{\rm gas}(R_{\rm cav,big-dust})/\Sigma_{\rm gas}(R_{\rm cav,gas})]} \,.
    \label{eq:w}
\end{equation}
As a point of clarification, the free parameters in the above
equations are $\delta_{\rm cav,gas}$, $\Sigma_{\rm 0,gas}$,
$R_{\rm c,gas}$, $\gamma_{\rm gas}$,
$R_{\rm cav,gas}$, and $R_{\rm cav,big-dust}$.
We connect the gas surface density to $R_{\rm cav,big-dust}$ so that
the gas pressure reaches a local maximum there --- see Figure \ref{fig:fiducial}.
This is motivated by dust filtration, which predicts
that mm-sized particles drift toward the pressure peak.
Inside $R_{\rm cav,gas}$, $\delta_{\rm cav,gas}=10^{-5}$ --- note
that this implies the gas surface density at 15 AU
is about $10^{-3}$ of the value at 70 AU.
The gas-to-dust ratio is 50:1 in the outer disk.
We note that our fiducial model overproduces the $^{13}$CO
and C$^{18}$O emission in the outer disk
(at the shortest baselines) compared to the data, but
we do not consider this discrepancy further as our focus in this
paper is on the inner cavity. 

We emphasize that the fiducial model does not provide a unique fit
to the data. With the exceptions of cavity radius and depth,
as discussed below, the constraints on many other
parameters are
rather weak
\citep[e.g.,][]{dong12pds70, dong12cavity, vandermarel15}.
Also, local non-axisymmetric features, such as the dip on the ring
at $H$-band, are not reproduced and are beyond the scope of this
paper.

\subsection{How Big are the Cavities?}\label{sec:size}

\begin{figure}
\begin{center}
\includegraphics[width=\textwidth]{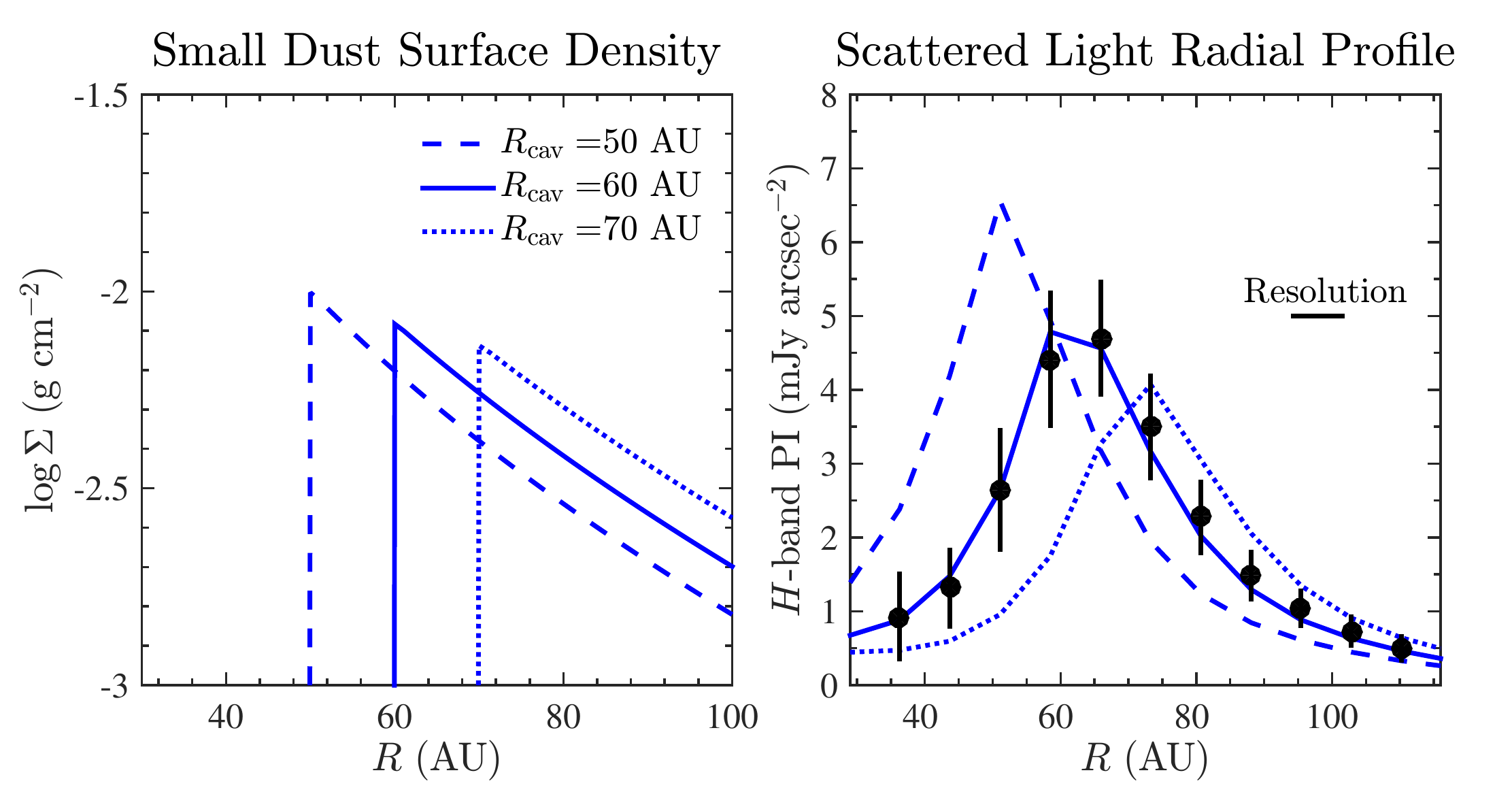}
\caption{The effect of varying $R_{\rm cav}$ in small dust on the scattered light image.}
\label{fig:nircavsize}
\end{center}
\end{figure}

\begin{figure}
\includegraphics[width=\textwidth]{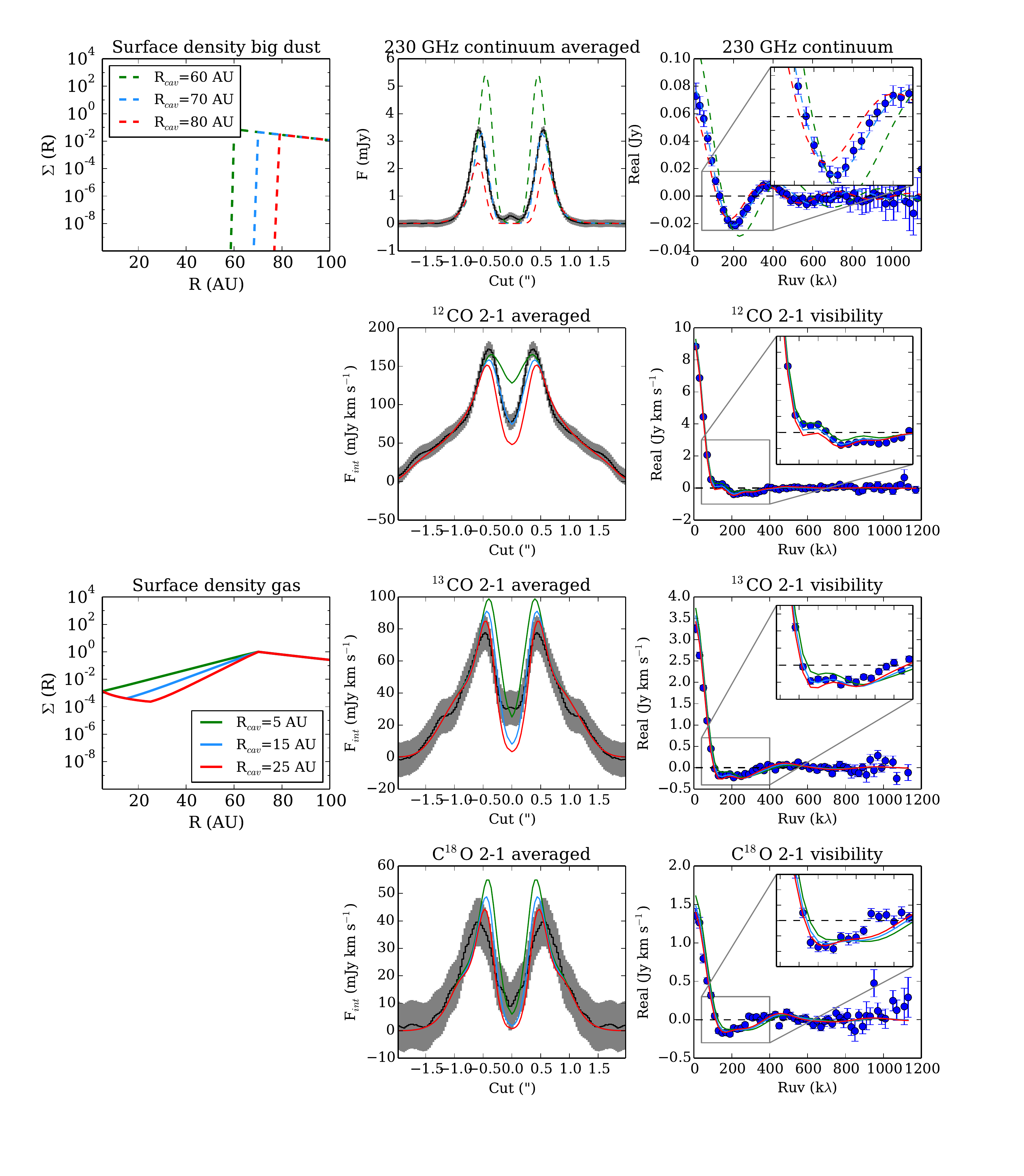}
\caption{Modeling results for different cavity sizes. The two panels in the left column show the surface density variations, the middle column shows the azimuthally averaged intensity cuts (the noise level is indicated by the gray zone; the model images have the same beam as the ALMA observations), and the right column shows the visibility profiles. {\bf Top to bottom in the middle and right columns}: ($1^{\rm st}$ row) the 230 GHz continuum, ($2^{\rm nd}$ row) the zero-moment maps of $^{12}$CO 2-1, ($3^{\rm rd}$ row) $^{13}$CO 2-1, and ($4^{\rm th}$ row) C$^{18}$O 2-1. The cavity sizes for the continuum are 60, 70 and 80 AU in 
green, blue, and red, respectively; 
and the gas cavity sizes are 5, 15 and 25 AU in green, blue, and red, respectively.}
\label{fig:mmcavsize}
\end{figure}

\begin{figure}
\includegraphics[width=0.75\textwidth]{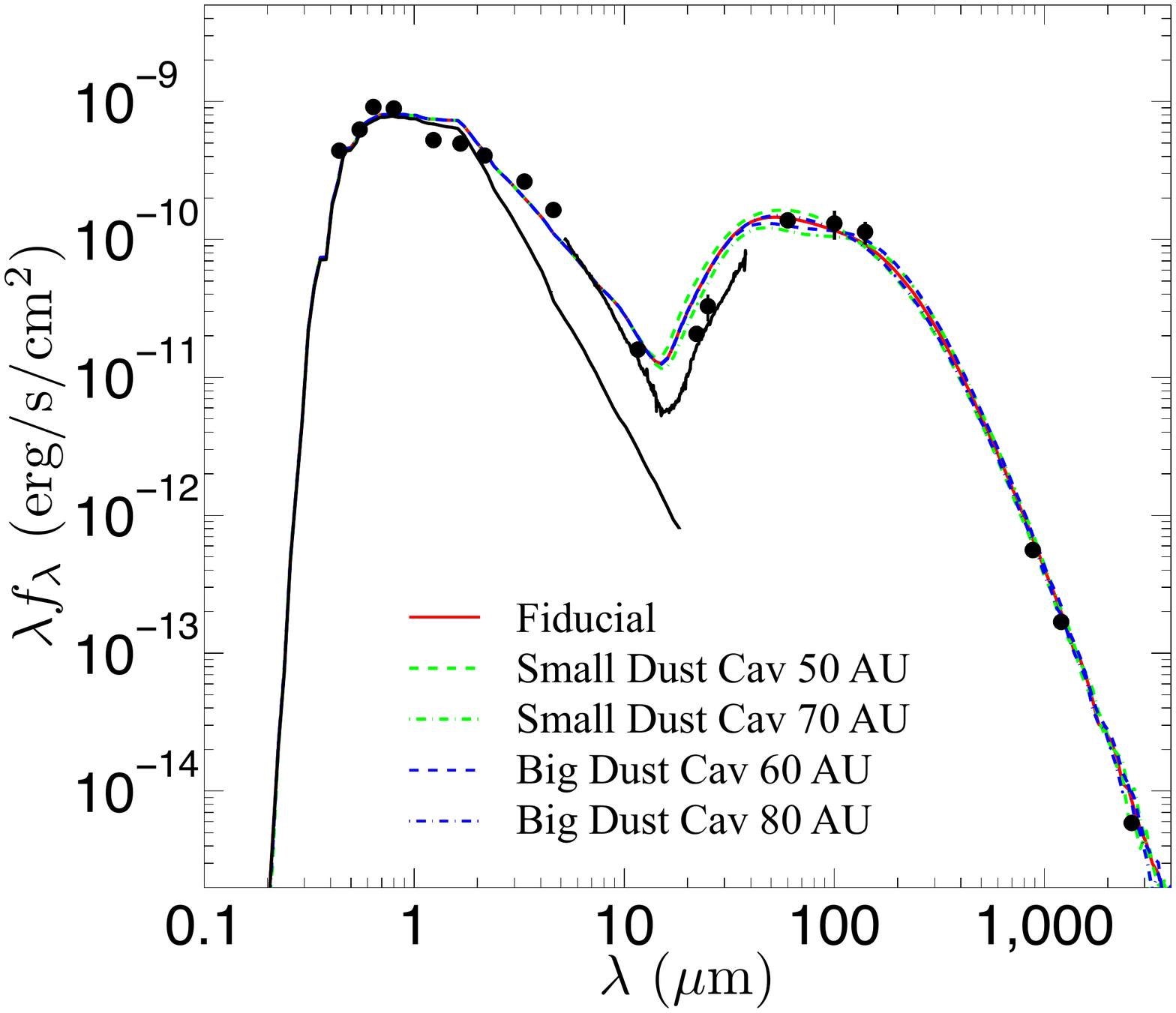}
\caption{The effect of varying the cavity sizes in the dust on the SED.}
\label{fig:sed_cavsize}
\end{figure}

To illustrate the effect of cavity size on various observations,
we show models with 3 cavity sizes for each disk component --- 50, 60, and 70 AU for $R_{\rm cav,small-dust}$ (Figure~\ref{fig:nircavsize}), 60, 70, and 80 AU for $R_{\rm cav,big-dust}$, and 5, 15, and 25 for $R_{\rm cav,gas}$ (Figure \ref{fig:mmcavsize}).  We focus on resolved observations as the SED only weakly depends on the cavity sizes within the range of our parameter exploration (Figure~\ref{fig:sed_cavsize}; the same is true for the discussions in Sections~\ref{sec:depletion} and \ref{sec:edge}). We observe the following:
\begin{itemize}
\item In scattered light, setting $R_{\rm cav,small-dust} = 50$ AU 
(70 AU) 
results in a cavity too small (too large) to be consistent with the
observations. The root-mean-square scatter of the observed $H$-band
radial profile can roughly tolerate a $\pm5$~AU deviation
from the fiducial model value of $R_{\rm cav,small-dust} = 60$ AU,
assuming a sharp cavity edge.
\item In the dust continuum, a smaller (bigger) cavity in the big
dust pushes the nulls on the visibility curve towards longer
(shorter) deprojected baselines.
The data are roughly consistent with $R_{\rm cav,big-dust} = 70 \pm 10$~AU.
\item In CO emission, the shape of the visibility profile
around the null changes noticeably when the cavity size is changed.
The cavity size in gas is roughly $R_{\rm cav,gas} = 15 \pm 10$ AU.
\end{itemize}
We conclude that the gas cavity is much smaller than the dust cavity,
while the cavity in the small dust is marginally smaller than in the
big dust.

\subsection{How Deep are the Cavities?}\label{sec:depletion}

\begin{figure}
\begin{center}
\includegraphics[width=\textwidth]{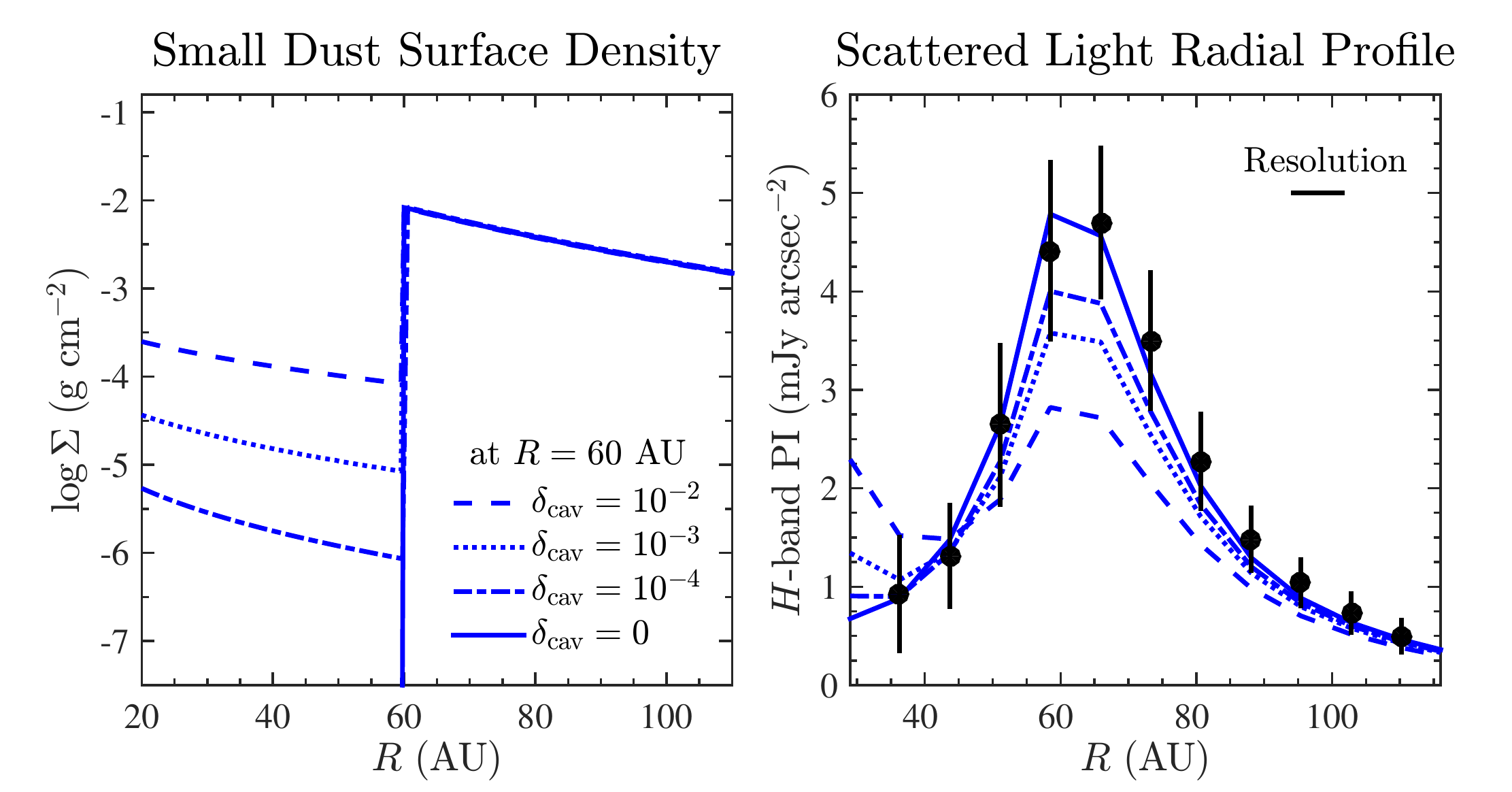}
\caption{The effect of varying $\delta_{\rm cav}$ in the small dust distribution.}
\label{fig:nircavdepth}
\end{center}
\end{figure}

Varying the cavity depletion has dramatic effects:

\begin{itemize}
\item In scattered light, increasing $\delta_{\rm cav,small-dust}$ smooths the ring and raises the surface brightness inside the cavity, as illustrated in Figure~\ref{fig:nircavdepth}. For these
extra models, $\Sigma_{\rm cav,small-dust}$ is depleted to $10^{-2}, 10^{-3}$, and $10^{-4}$ of its value at the cavity edge, and smoothly joins the fixed rim at 0.07 AU with $\delta_{\rm cav,small-dust}=10^{-6}$ in order to match the SED; at $R>10$ AU, $\delta_{\rm cav}$ is nearly constant, i.e.,
$\delta_{\rm cav,small-dust}(R)\approx\delta_{\rm cav,small-dust}$($R=$ 60 AU).
To be consistent within error bars with the Subaru data---in
particular to reproduce the contrast of the cavity---the drop in
small dust surface density beyond 10 AU needs to be
at least a factor of 1000;
indeed the data are consistent with no small dust at all, as in the
fiducial model.

\item  In the millimeter dust continuum, a depletion in the big dust of less than 1000 results in excess emission in the center of the image and a vertical offset in the visibility curve. The data are consistent with no big dust at all. Thus we conclude $\delta_{\rm cav,big-dust}\leq10^{-3}$.

\item In CO emission, the visibility data and null positions
appear to constrain $\delta_{\rm cav,gas}$ to within a
factor of 10 of our fiducial value.

\end{itemize}

We conclude that the gas cavity is shallower than the small-dust
and big-dust cavities. The constraints on the cavity size and depth are summarized in Table~\ref{tab:cavityproperties} (assuming fiducial edge sharpnesses).

\subsection{How Sharp are the Cavity Edges?}\label{sec:edge}
 
In this section, we explore ``smooth'' cavity edge structures in 
small and big dust, for
which the transition from the cavity region to
the outer disk occurs over a finite radius range. This profile
is motivated by the shapes of the cavities in planet-disk interaction
models \citep[e.g.,][]{duffell15gap, fung16}. A key question is
whether the big and small dust grains might have cavity
edges having different shapes. There are many ways to model smooth
cavity edges, and we restrict the discussion to a few illustrative
cases. For the gas distribution, we show the effect of having a
sharp cavity edge, and show that a smooth cavity edge is demanded by
the CO observations.

\begin{figure}
\begin{center}
\includegraphics[width=\textwidth]{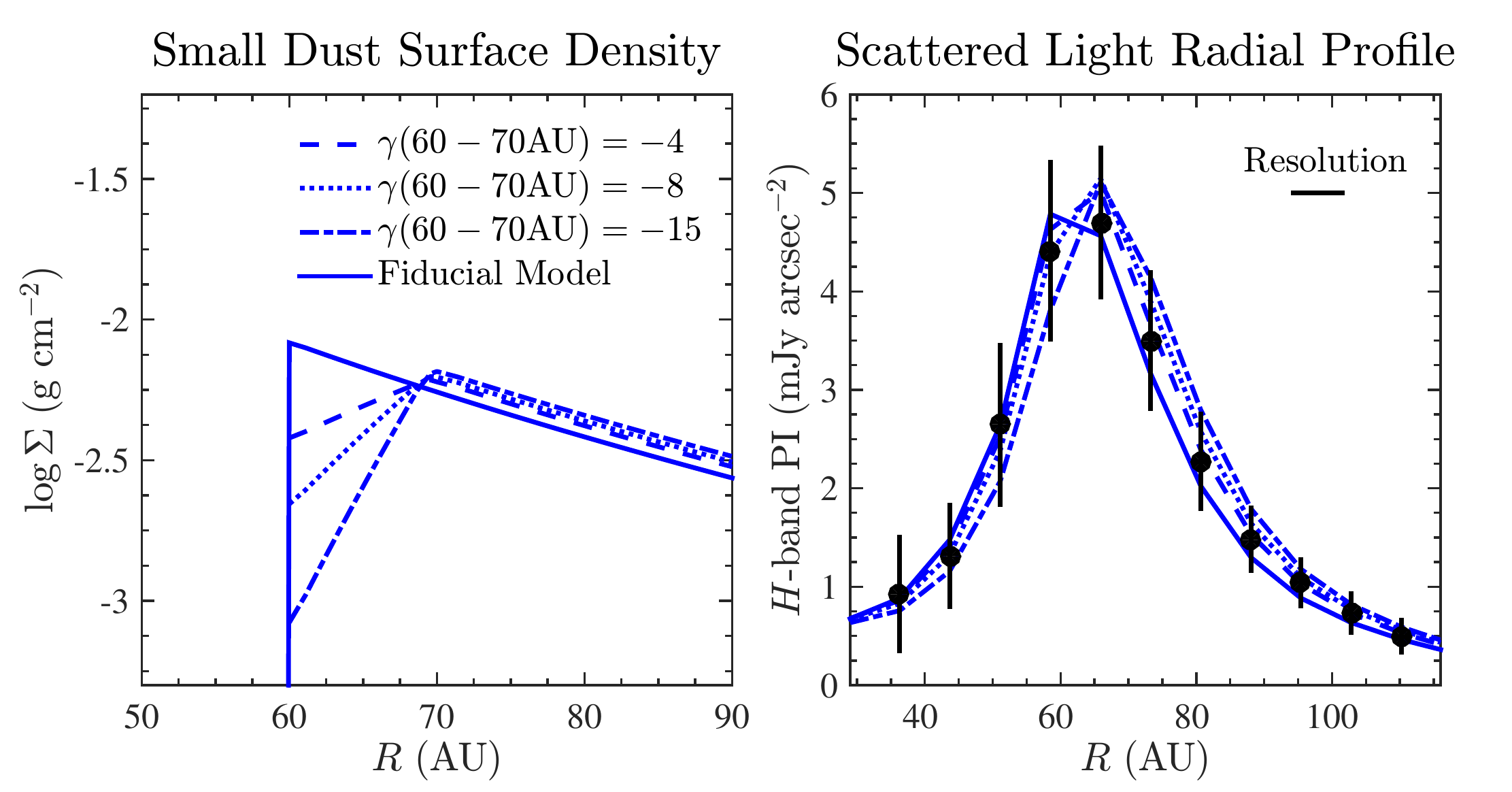}
\caption{The effect of a smooth cavity edge transition between 60 AU and 70 AU in the small dust on the scattered light image.}
\label{fig:nircavedge}
\end{center}
\end{figure}

\begin{figure}
\includegraphics[width=\textwidth]{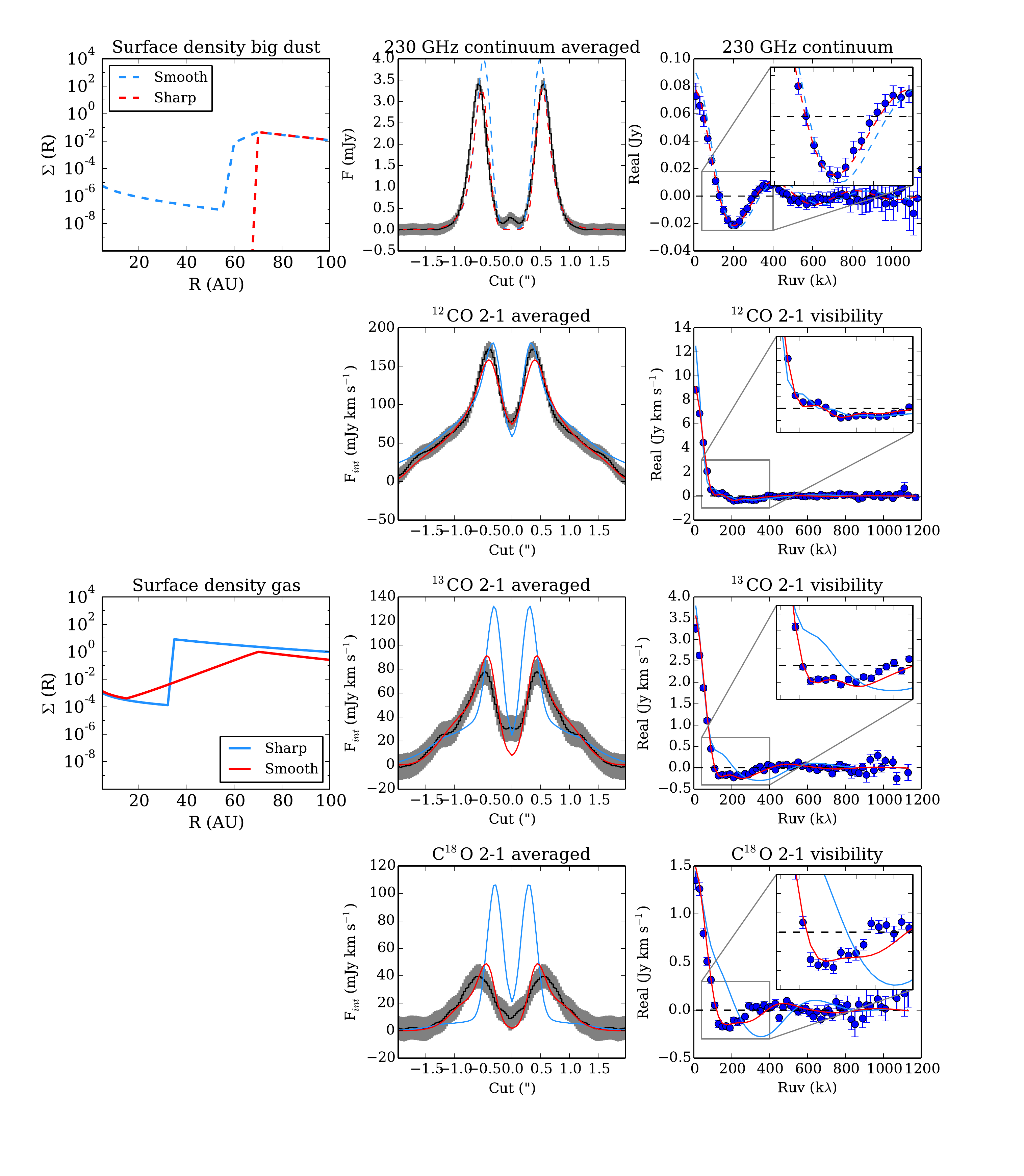}
\caption{Modeling results for sharp vs.~smooth density drops in gas and in big dust (such as used in \citealt{vandermarel15}) compared with observations. The two panels in the left column show the surface density variations, the middle column shows the azimuthally averaged intensity cuts (the noise level is indicated by the gray zone; the model images have the same beam as the ALMA observations), and the right column shows the visibility profiles. {\bf Top to bottom in the middle and right columns}: ($1^{\rm st}$ row) the 230 GHz continuum, ($2^{\rm nd}$ row) the zero-moment maps of $^{12}$CO 2-1, ($3^{\rm rd}$ row) $^{13}$CO 2-1, and ($4^{\rm th}$ row) C$^{18}$O 2-1.}
\label{fig:mmcavedge}
\end{figure}

\begin{itemize}
\item Figure~\ref{fig:nircavedge} shows three models for the
scattered light where transition regions from 60 to 70 AU are
constructed to join the inner and outer disk in small dust. 
Introducing these smooth structures widens the cavity edge in scattered light. In addition, the ring shifts outward with a more abrupt transition at 70 AU and a smoother transition at 60 AU, as the 70 AU break gradually becomes the ``new'' cavity edge; the structures in scattered light trace the most abrupt changes in dust surface density. Overall, all the models considered
appear consistent with the data.
\item Figure \ref{fig:mmcavedge} shows that a smooth drop in the
big-dust distribution shifts the peak in the mm continuum emission
toward smaller radius, and the model does not match the data at large
baselines. Thus a smooth cavity edge in big dust appears
inconsistent with the data.
\item In gas observations, \citet{vandermarel15} successfully fitted 
the ALMA Cycle 0 $^{12}$CO $J=$ 3-2 observations with a sharp gas
cavity edge at 30 AU; however, this model cannot fit the new ALMA CO
data. As shown in Figure~\ref{fig:mmcavedge}, a sharp drop-off at
$R_{\rm cav,gas}=30$ AU gives model visibility profiles of the other
CO isotope lines that have clear deviations around the location of
the null (although the total integrated flux is similar and the 
$^{12}$CO 2-1 visibility profile fits reasonably well).
Also, a double-drop model (two sharp $\Sigma_{\rm gas}$ drops,
one at $R_{\rm cav,big-dust}$, the other at $R_{\rm cav,gas}$),
such as proposed by \citet{vandermarel16},
is not able to reproduce the visibility profiles
of the CO isotope lines. 
\end{itemize}

We conclude that the current data in scattered light cannot
distinguish a sharp cavity edge from a
smooth transition that takes places over 10 AU
in the vicinity of the cavity edge, whereas
the mm continuum data appear to require a
relatively sharp cavity edge.
The cavity edge structures in big and small dust
can be approximately co-spatial.
By comparison, the CO line data demand a smooth
transition in gas density between 15 and 70 AU. Models with smooth
gas cavity edges as in the fiducial model were also fitted
successfully to previous CO isotopologue data of other transitional
disks at lower spatial resolution \citep{vandermarel16}.


\section{Summary and Discussion}\label{sec:summary}
 
In this paper, we report on ALMA Cycle 2 observations of the
transitional disk J160421.7-213028 (J1640)
in 230 GHz continuum and $^{12}$CO,
$^{13}$CO, and C$^{18}$O $J=$2--1 emission. Using radiative transfer
simulations, we construct a simple disk+cavity model
to account for the
spectral energy distribution (SED), and for resolved observations
of the system in near-infrared (NIR) scattered light, dust continuum
emission, and CO line emission. We constrain the radius of the
cavity, its depth, and the sharpness of its edge.
Our main results are as follows:
\begin{enumerate}
\item Our fiducial model, which fits the observations adequately by eye, has a completely empty cavity in both sub-$\micron$-sized small dust and mm-sized big dust,\footnote{Apart from a possible sub-0.1~AU dusty inner rim needed to account for a variable NIR excess.} of radius of 60--70 AU. The gas exhibits a cavity 15 AU in radius that is uniformly depleted by a factor of $10^5$; the gas surface density grades smoothly from the edge of this gas cavity to the outer (undepleted) disk at 70 AU.
\item The NIR scattered light observations constrain the cavity radius in small dust to within $\pm$5 AU from the best-fit value of 60 AU. The first null on the visibility curves of the dust continuum and C$^{18}$O emission constrain the cavity size in the big dust and in the gas to within $\pm$10 AU of 70 AU and 15 AU, respectively. Thus the data are consistent with the same cavity size in the two dust populations, while both dust cavities are significantly larger than the gas cavity.
\item While the scattered light and mm continuum data are consistent with a completely empty cavity in both  small and big dust, the data can tolerate a finite factor of $10^3$ depletion in both. The CO line data require 
gas surface densities at 15 AU to be depleted by factors of $10^2$--$10^4$ relative to gas surface densities at 70 AU.
\item Currently, the NIR data cannot distinguish between a sharp cavity edge in the small dust population and a transition region of annular width 10 AU. The CO observations demand that gas densities vary smoothly inside 70 AU; a sharp gas cavity edge at 30 AU, or a model that drops in gas surface density at both 15 AU and 70 AU, cannot fit the data.
\end{enumerate}

{\it What is the nature of J1604's cavity?} J1604 bears a number of characteristics common to other transitional disks. It shows clear signs of a depleted inner region in all disk components, with a gas cavity that is substantially smaller than the dust cavity. 
Similar structures have been seen in a few transitional disks \citep{vandermarel15, vandermarel16},
and are interpreted as evidence of
dynamical clearing by (multiple) planets inside the cavities,
coupled with dust filtration
\citep[e.g.,][]{pinilla12-diffcavsize, zhu12}.

On the other hand, J1604 is unique among its peers in several ways. While all the other sources in the van der Marel ALMA CO disk sample have accretion rates on the order of 10$^{-9}M_\odot$ yr$^{-1}$, J1604 has little accretion, no larger than $10^{-11}M_\odot$ yr$^{-1}$ \citep{dahm09, dahm12, mathews12}. Furthermore, its innermost dust disk of radius $\sim0.1$~AU may be time variable, with IRAC data indicating no NIR excess and the WISE photometry and Spitzer IRS spectrum indicating a small NIR excess (Table~\ref{tab:sed}). In addition, while many transitional disks appear to be asymmetric in the ALMA dust continuum \citep[e.g.,][]{vandermarel13, casassus13, perez14}, J1604 has a nearly symmetric ring
in thermal emission. At the same time, scattered light imaging shows that the ring has a small gap of time-varying position angle along its circumference \citep{mayama12, pinilla15j1604} --- see Figure \ref{fig:nir_fiducial}.

As discussed in Section~\ref{sec:intro}, both dynamical sculpting by planets and photoevaporation can open cavities in disks. The main observational distinction between the two is whether the system still has significant accretion onto the central star. Transitional disks with significant accretion (on the order of $10^{-9}M_\odot$ yr$^{-1}$) may be compatible with planets, while those with negligible accretion may indicate starvation by photoevaporation \citep[e.g.,][]{alexander09, cieza12, owen12clarke, espaillat14}. The observed small-to-zero accretion rate in J1604 suggests the latter. However, the smooth gas gap edge transition in J1604 appears inconsistent with photoevaporation (Section~\ref{sec:intro}; e.g., \citealt{alexander07}), even as the small dust grains---which are expected to be well coupled to the gas---do exhibit a relatively sharp drop in density at the cavity edge. The difference between the small dust and gas edge structures is intriguing and merits future study. Finally, we note that if planets are responsible for opening the cavity, the size and depth of the cavity imply a chain of multiple massive planets, likely of multi-Jupiter masses, unless the disk viscosity $\alpha$ is substantially lower than $10^{-3}$ \citep{fung14, dong16td}. Such planets at large radius have been found to be rare \citep{bowler16}.

There are at least two paths forward for exploring the origin of the cavity in J1604. First, if the cavity is opened by (multiple) giant planets, these may be detected in deep direct imaging observations. So far the deepest exposure for J1604 is by \citet{canovas16}, in which no planet candidates were found, and objects of mass 2--3 $\mj$  outside 25 AU have been ruled out according to the hot-start planet formation model. Second, if the disk is currently being photoevaporated, blueshifted optical and infrared forbidden lines indicating a photoevaporative wind may be detectable in this nearly face-on system (see the example in TW Hya, \citealt{pascucci11}). Further observations are needed to elucidate the nature of J1604's cavities.


\section*{Acknowledgments}
We are grateful to the anonymous referee for constructive suggestions that improved the quality of the paper. This project is partially supported by NASA through Hubble Fellowship grant HST-HF-51320.01-A (R.D.) awarded by the Space Telescope Science Institute, which is operated by the Association of Universities for Research in Astronomy, Inc., for NASA, under contract NAS 5-26555. Numerical calculations were performed on the SAVIO cluster provided by the Berkeley Research Computing program, supported by the UC Berkeley Vice Chancellor for Research and the Berkeley Center for Integrative Planetary Science. This paper makes use of the following ALMA data: ADS/JAO.ALMA\#2013.1.01020S. ALMA is a partnership of ESO (representing its member states), NSF (USA) and NINS (Japan), together with NRC (Canada), NSC and ASIAA (Taiwan), and KASI (Republic of Korea), in cooperation with the Republic of Chile. The Joint ALMA Observatory is operated by ESO, AUI/NRAO and NAOJ. The National Radio Astronomy Observatory is a facility of the National Science Foundation operated under cooperative agreement by Associated Universities, Inc.


\clearpage

\begin{table}[]
    \centering
    \begin{tabular}{llllll}
         \hline
         \hline
         Line&Rest&Beam&&rms\tablenotemark{(a)}&peak\tablenotemark{(a)}  \\
         &frequency (GHz)&size($\arcsec$)&PA($^{\circ}$)&(mJy/beam)&(mJy/beam) \\
         \hline
         $^{12}$CO 2~$\rightarrow$~1&230.53800&0.27$\times$0.23&59&10&226\\
         $^{13}$CO 2~$\rightarrow$~1&220.39868&0.28$\times$0.24&58&11&168\\
         C$^{18}$O 2~$\rightarrow$~1&219.56036&0.28$\times$0.24&58&11&106\\
         Continuum&234.00000&0.24$\times$0.21&61&0.11&4.0\\
         \hline
         \hline
    \end{tabular}
    \caption{Properties of the ALMA observations. $^{(a)}$ Measured in 0.25 km/s bins.}
    \label{tab:obsproperties}
\end{table}


\begin{table}[]
\centering
\begin{tabular}{lccc}
\hline
                   & Gas & Small Dust & Big Dust \\ \hline
Total Mass ($\mj$)       & 2.5    &     0.013       &   0.066       \\
$R_{\rm c}$ (AU)       &  30   &      100     &   30       \\
$R_{\rm out}$ (AU)     &  300   &     200      &   300       \\
$\gamma$           &   1.0     &      2.0      &    1.0      \\
$\psi$             &   0.5  &     1.1       &   0.5       \\
\hline
$R_{\rm cav}$ (AU)      &  15\tablenotemark{(a)}  &      60       &    70      \\
$\delta_{\rm cav}$ &  $10^{-5}$   &  0        &    0      \\
\hline
\end{tabular}
\caption{Parameters of the fiducial model. The last two parameters ($R_{\rm cav}$ and $\delta_{\rm cav}$) are varied in Section~\ref{sec:properties}. $^{(a)}$ The gas surface density starts to decrease at 70 AU and reaches a minimum at 15 AU --- see equation (\ref{eq:sigmagas}) and Figure \ref{fig:fiducial}.}
\label{tab:base}
\end{table}


  \begin{table}
    \begin{center}
      \caption{Archival SED data for J160421.7-213028}
      \begin{tabular}{lcc} 
        \tableline
        \tableline
	Wavelength                         & $F_{\nu}$ (mJy)    & Note           \\
	                                   &                    &                \\
	\tableline
	$B$\tablenotemark{a,b}             &    64.6            & \citet{preibisch99} \\
	$V$\tablenotemark{a,b}             &   114.9            & \citet{zacharias05} \\
	$R$\tablenotemark{a,b}             &   194.8            & \citet{preibisch99} \\
	$I$\tablenotemark{a,b}             &   237.6            & \citet{preibisch99} \\
	2MASS ($J$)\tablenotemark{a,b}     &   216.0 $\pm$  4.6 & \citet{cutri03} \\
	2MASS ($H$)\tablenotemark{a,b}     &   275.0 $\pm$  6.1 & \citet{cutri03} \\
	2MASS ($Ks$)\tablenotemark{a,b}    &   292.0 $\pm$  5.6 & \citet{cutri03} \\
	WISE (3.4 $\mu$m)\tablenotemark{b} &   293.4 $\pm$  5.9 & \citet{cutri12} \\
	WISE (4.6 $\mu$m)\tablenotemark{b} &   251.0 $\pm$  4.2 & \citet{cutri12} \\
	WISE (12 $\mu$m)\tablenotemark{b}  &    61.5 $\pm$  0.8 & \citet{cutri12} \\
	WISE (22 $\mu$m)\tablenotemark{b}  &   152.1 $\pm$  3.4 & \citet{cutri12} \\
	IRAC (4.5 $\mu$m)                  &    62.7 $\pm$  0.8 & \citet{carpenter06}  \\
	IRAC (8.0 $\mu$m)                  &    26.3 $\pm$  0.2 & \citet{carpenter06}  \\
	IRAC (16.0 $\mu$m)                 &    26.8 $\pm$  0.2 & \citet{carpenter06}  \\
	IRAS (25 $\mu$m)                   &   273.2 $\pm$ 60.1 & \citet{moshir89} \\
	IRAS (60 $\mu$m)                   &  2754   $\pm$ 170.7 & \citet{moshir89} \\
	IRAS (100 $\mu$m)                  &  4355   $\pm$ 1045.2 & \citet{moshir89} \\
	AKARI (140 $\mu$m)                 &  5288.7 $\pm$ 1000.0 & VizieR II/298  \\
	880 $\mu$m                         &   164   $\pm$ 6      & \citet{mathews12} \\
	1.2 mm                             &    67.5 $\pm$ 1.4    & \citet{mathews12} \\
	2.6 mm                             &     5.1 $\pm$ 0.5    & \citet{mathews12} \\
	IRS                                &   --- & Spitzer Heritage Archive\tablenotemark{c}  \\
	\tableline
      \end{tabular}
      \tablenotetext{a}{
        The extinction law was adopted from \citet{mathis90}, assuming $A_v=1$ for J1604 \citep{preibisch02}.
      }
      \tablenotetext{b}{
        Absolute flux conversions in optical, 2MASS, WISE photometric data were adopted
          from \citet{bessell98}, \citet{cohen03}, and \citet{jarrett11}, respectively.
      }
      \tablenotetext{c}{
	This work is based in part on observations made with the Spitzer Space Telescope, obtained from the NASA/ IPAC Infrared Science Archive, both of which are operated by the Jet Propulsion Laboratory, California Institute of Technology under a contract with the National Aeronautics and Space Administration.
      }
    \label{tab:sed}
    \end{center}
  \end{table}  
  
\begin{table}[]
\centering
\begin{tabular}{l|c|c}
\toprule
Component    & $R_{\rm cav}$ (AU)\ \ \  & $\delta_{\rm cav}$ \\ \Xhline{2\arrayrulewidth}
Small Dust &   60$\pm$5            &     $\lesssim10^{-3}$               \\ \hline
Big Dust   &   70$\pm$10            &      $\lesssim10^{-3}$                    \\ \hline
Gas        & 15$\pm$10               &      $10^{-6}\sim10^{-4}$               \\ \bottomrule
\end{tabular}
\caption{Constraints on the cavity depth and radius in the three disk components assuming the cavity edge profiles are as in the fiducial model. See Section~\ref{sec:properties} for details.}
\label{tab:cavityproperties}
\end{table}


\begin{figure}
\includegraphics[width=\textwidth]{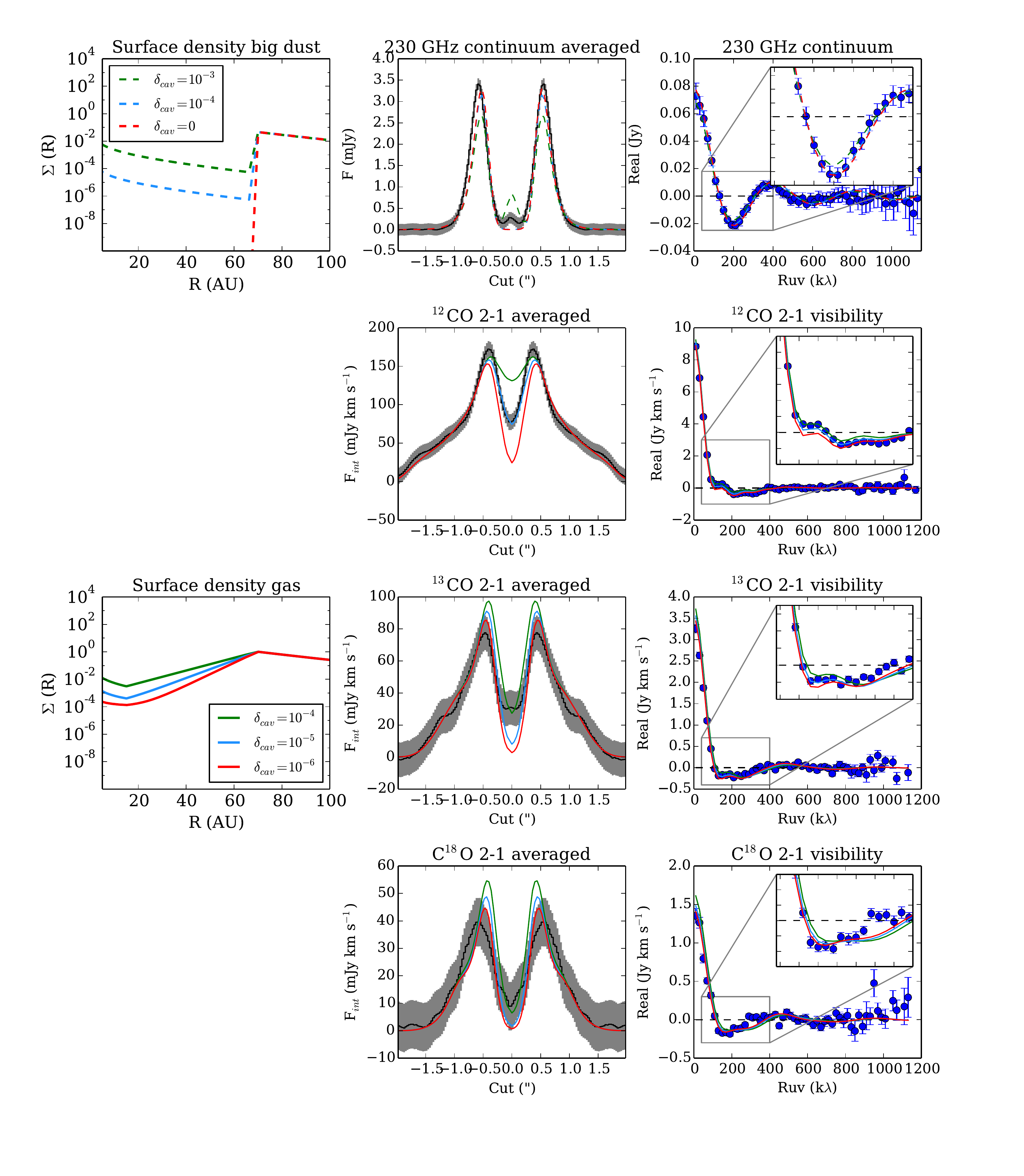}
\caption{Modeling results for different cavity depths in the gas density. The two panels in the left column show the surface density variations, the middle column shows the azimuthally averaged intensity cuts (the noise level is indicated by the gray zone; the model images have the same beam as the ALMA observations), and the right column shows the visibility profiles. {\bf Top to bottom in the middle and right columns}: ($1^{\rm st}$ row) the 230 GHz continuum, 
($2^{\rm nd}$ row) the zero-moment maps of $^{12}$CO 2-1, ($3^{\rm rd}$ row) $^{13}$CO 2-1, and ($4^{\rm th}$ row) C$^{18}$O 2-1, with depths of 10$^{-4}$, 10$^{-5}$, and 10$^{-6}$. In the dust surface density, green, blue, red curves are for 10$^{-3}$, 10$^{-4}$, and infinite depletion, respectively.}
\label{fig:mmcavitydepth}
\end{figure}

\clearpage

\end{document}